\documentclass[journal=jpclcd,manuscript=article]{achemso}

\usepackage{url}
\usepackage{chemformula} 
\usepackage[T1]{fontenc} 
\author{Khadijeh Najafi}
\affiliation[Unknown University]
{Department of Physics, Virginia Tech, Blacksburg, VA 24061, U.S.A}
\email{najafi.khadijeh@gmail.com}

\author{Aleksander L. Wysocki}
\affiliation[Unknown University]
{Department of Physics, Virginia Tech, Blacksburg, VA 24061, U.S.A}

\author{Kyungwha Park}
\affiliation[Unknown University]
{Department of Physics, Virginia Tech, Blacksburg, VA 24061, U.S.A}

\author{Sophia E. Economou}
\affiliation[Unknown University]
{Department of Physics, Virginia Tech, Blacksburg, VA 24061, U.S.A}

\author{Edwin Barnes}
\affiliation[Unknown University]
{Department of Physics, Virginia Tech, Blacksburg, VA 24061, U.S.A}

\title[An \textsf{achemso} demo]{Toward Long-Range Entanglement Between Electrically Driven Single-Molecule Magnets}

\abbreviations{IR,NMR,UV}

\begin{document}

\begin{tocentry}

\includegraphics[width=\columnwidth]{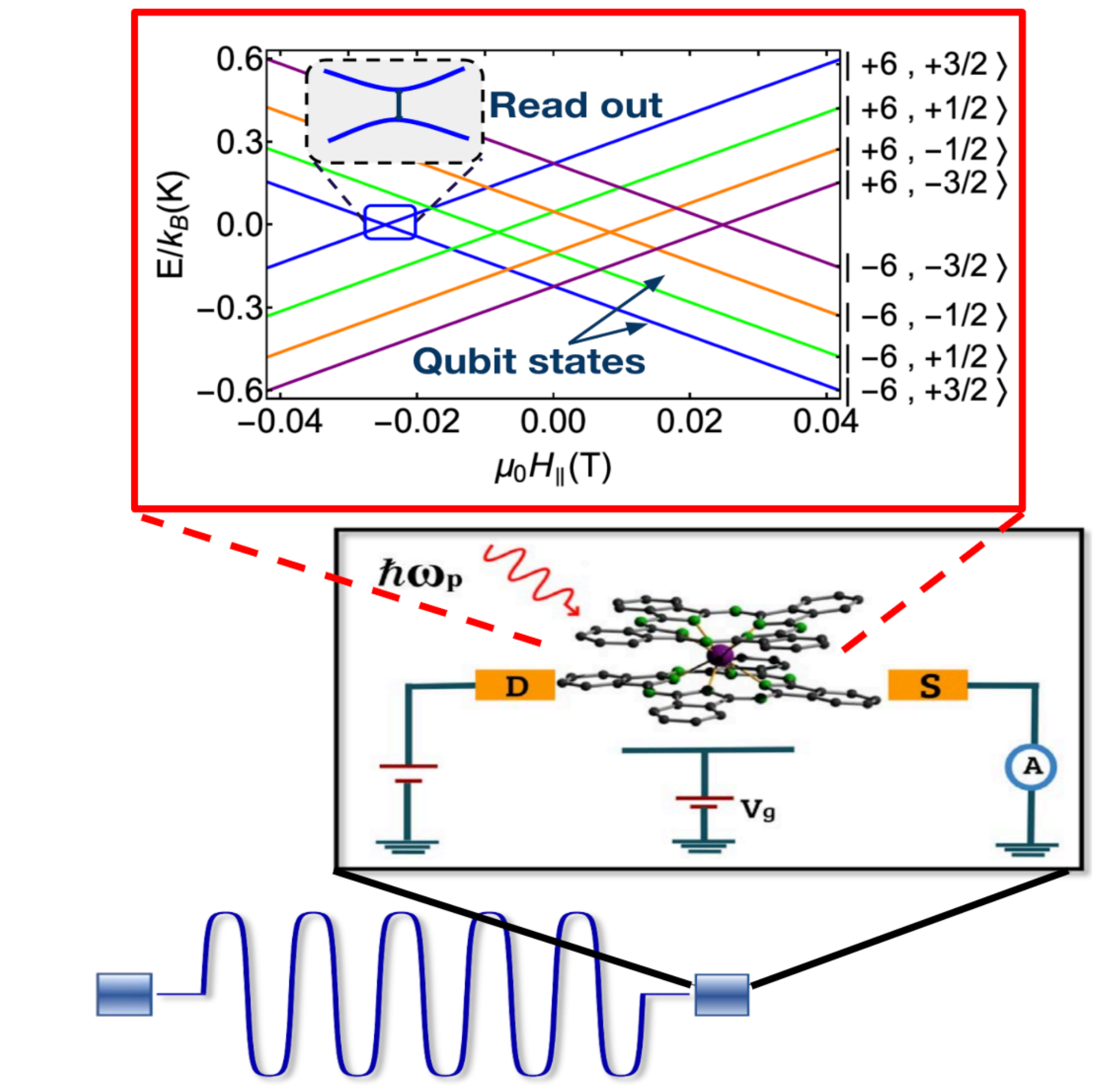}





\end{tocentry}

\begin{abstract}
Over the past two decades, several molecules have been explored as possible building blocks of a quantum computer, a device that would provide exponential speedups for a number of problems, including the simulation of large, strongly correlated chemical systems. Achieving strong interactions and entanglement between molecular qubits remains an outstanding challenge. Here, we show that the TbPc$_2$ single-molecule magnet has the potential to overcome this obstacle due to its sensitivity to electric fields stemming from the hyperfine Stark effect. We show how this feature can be leveraged to achieve long-range entanglement between pairs of molecules using a superconducting resonator as a mediator. Our results suggest that the molecule-resonator interaction is near the edge of the strong-coupling regime and could potentially pass into it given a more detailed, quantitative understanding of the TbPc$_2$ molecule.
\end{abstract}


Quantum information processing (QIP) is based on storing information in quantum two-level systems (qubits) and takes full advantage of key features of quantum mechanics, such as quantum interference and entanglement, in order to exponentially speed up certain types of problems.\cite{Nielsen_Chuang} The most well-known example  is Shor's algorithm,\cite{Shor_SIAM97} which, if implemented in a large quantum computer, would be able to break the RSA cryptosystem, which is currently the predominant system for securely transmitting information. While the requirements for Shor's algorithm are formidable in terms of the necessary number of qubits and level of control precision, there also exist important near-term applications that can be implemented with a more modestly sized quantum computer. A notable application is quantum simulation,\cite{Lanyon2010,Toloui2013,Sugisaki2016} which would enable the computational modeling of large-scale strongly correlated quantum systems,
with applications in quantum chemistry and medicine.\cite{Sugisaki2016}

Over the past two decades, several quantum systems have been explored as candidate qubits for QIP. An obvious choice is spin (either electronic or nuclear), as it can be a true two-level system and tends to be well isolated from its environment, leading to relatively long coherence times.\cite{Muhonen_NatNano14,Abobeih_NatCommun18}
In 2001, Leuenberger and Loss proposed to use the spin of the nanoscale single-molecule magnet (SMM) Mn$_{12}$ as a qubit, with the control achieved via electron spin resonance pulses. \cite{LEUE01,Tejada2001} Since then, plausible setups and architectures for quantum computing with SMMs have been proposed
by several groups. \cite{Schlegel2008,Takahashi2011,Martinez2012,Luis2016,Hill2016,Riaz_JACS}
A majority of the proposals are based on magnetically controlled SMM {\it electron} spin
qubits, for which the coherence times are not yet sufficiently long for quantum computing. Magnetic field control also limits the potential of SMMs
for device integration and scalability, as it is extremely challenging to address individual qubits this way, and it also tends to yield slow gate operations, limiting the complexity of algorithms that can be run.

Recently, a qubit candidate with remarkable properties was experimentally demonstrated by Thiele et al.:\cite{Thiele_Science14} the SMM TbPc$_2$, which features a \emph{nuclear} spin as the qubit, with the attractive and unusual property of being electrically controllable. This combines the best of both worlds: long-lived qubit coherence with fast controllability. This recent exciting discovery opens up the opportunity for the development of scalable SMM-based QIP devices.

To develop a quantum information processor, the qubits must fulfill certain criteria. First of all, individual qubits must be controllable and measurable. Proof-of-principle demonstrations of these capabilities have been carried out for TbPc$_2$ SMMs.\cite{Thiele_Science14} A second crucial requirement for QIP is that the qubits must be coupled to each other through some physical interaction in order to implement quantum logic gates. The direct dipolar coupling between nuclear spins is far too weak to achieve significant coupling. Most proposals instead posit using electron spins as mediators of an effective nuclear spin coupling.\cite{Kane_Nature98,Pla_Nature13,Kalra_PRX14} However, electron spin dipolar interactions are also weak, and exchange coupling requires the daunting task of placing donors or molecules with nanometer precision.

\begin{figure} [h] 
\includegraphics[width=1.\columnwidth]{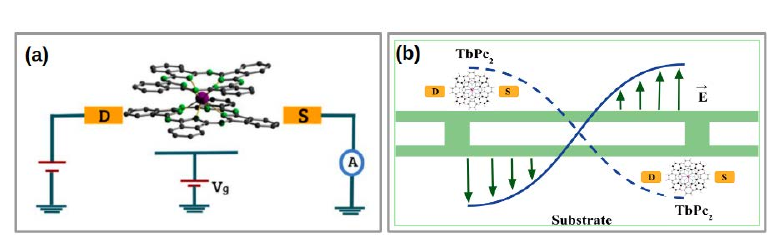}
\caption{ Schematic of TbPc$_2$ single-molecule transistors coupled to a resonator. (a) Each molecule is comprised of a Tb ion sandwiched between two flat Pc ligands that are connected to a source and drain. Individual molecules are driven via the hyperfine Stark effect by modulating the voltage across the source and drain. (b) Two molecules coupled to the electric field of a common cavity resonator via the hyperfine Stark effect. The molecules are located at the antinodes of the cavity's electric field.} 
\label{fig:smm_resonator}
\end{figure}

A possible way to overcome these challenges is to use a superconducting transmission line resonator as a `bus' to mediate coupling between TbPc$_2$ SMMs. The fact that TbPc$_2$ SMMs are sensitive to electric fields through the hyperfine Stark effect\cite{Thiele_Science14} allows for them to couple to the electric field of the resonator, potentially leading to strong interqubit interactions. This approach is also natural given that the splittings between nuclear spin states in a TbPc$_2$ molecule are on the order of GHz, the typical frequency range of superconducting resonators.\cite{Day_Nature03,Wallraff_Nature04}
Furthermore, this method of coupling has the advantage that it is long-range, enabling direct coupling between pairs of distant qubits, something not possible with nearest-neighbor architectures. Superconducting resonators are widely used to couple qubits based on superconducting circuits,\cite{Majer_Nature07,Steffen_IBM11,Corcoles_NC15} and have also been employed to couple remote electron spins.\cite{Schuster_PRL10,Mi_Science18,Borjans_arxiv19,Burkard_arxiv19,Landig_Nature18,Landig_arxiv19,Samkharadze_Science18} They have high quality factors ($Q\sim10^6$)\cite{Megrant_APL12,Bruno_APL15} and mature fabrication technology. Moreover, the flat structure of the TbPc$_2$ molecule (Fig.~\ref{fig:smm_resonator}) makes the prospects for fabrication with a superconducting resonator promising. A schematic of the envisioned architecture is shown for two molecules in Fig.~\ref{fig:smm_resonator}.

In this paper, we investigate the feasibility of using superconducting resonators to create entanglement between TbPc$_2$ qubits. We begin by modeling the single-qubit Rabi oscillations observed in Ref.~\cite{Thiele_Science14} to determine the minimal hyperfine tensor necessary to reproduce these findings. We use this information to estimate the qubit-resonator coupling, which we find is on the edge of the strong coupling regime defined by the cavity photon loss rate and the spin dephasing time. We then employ these results to construct an effective Hamiltonian for multiple TbPc$_2$ qubits coupled to the electric field of a superconducting resonator via the hyperfine Stark effect. To test the entangling capabilities of this effective interaction, we design two-qubit CNOT gates and determine the fidelities and gate speeds for a range of coupling strengths. We find that while fidelities above 99\% can be achieved in all cases, pushing gate times to well above the qubit dephasing time will likely require boosting the interaction strength further. Our results suggest that superconducting resonators may be a promising approach for building quantum processors out of electrically driven SMMs, although further improvements in device designs will likely be needed to reach the strong coupling regime.

\begin{figure} [h] 
\includegraphics[width=0.8\columnwidth]{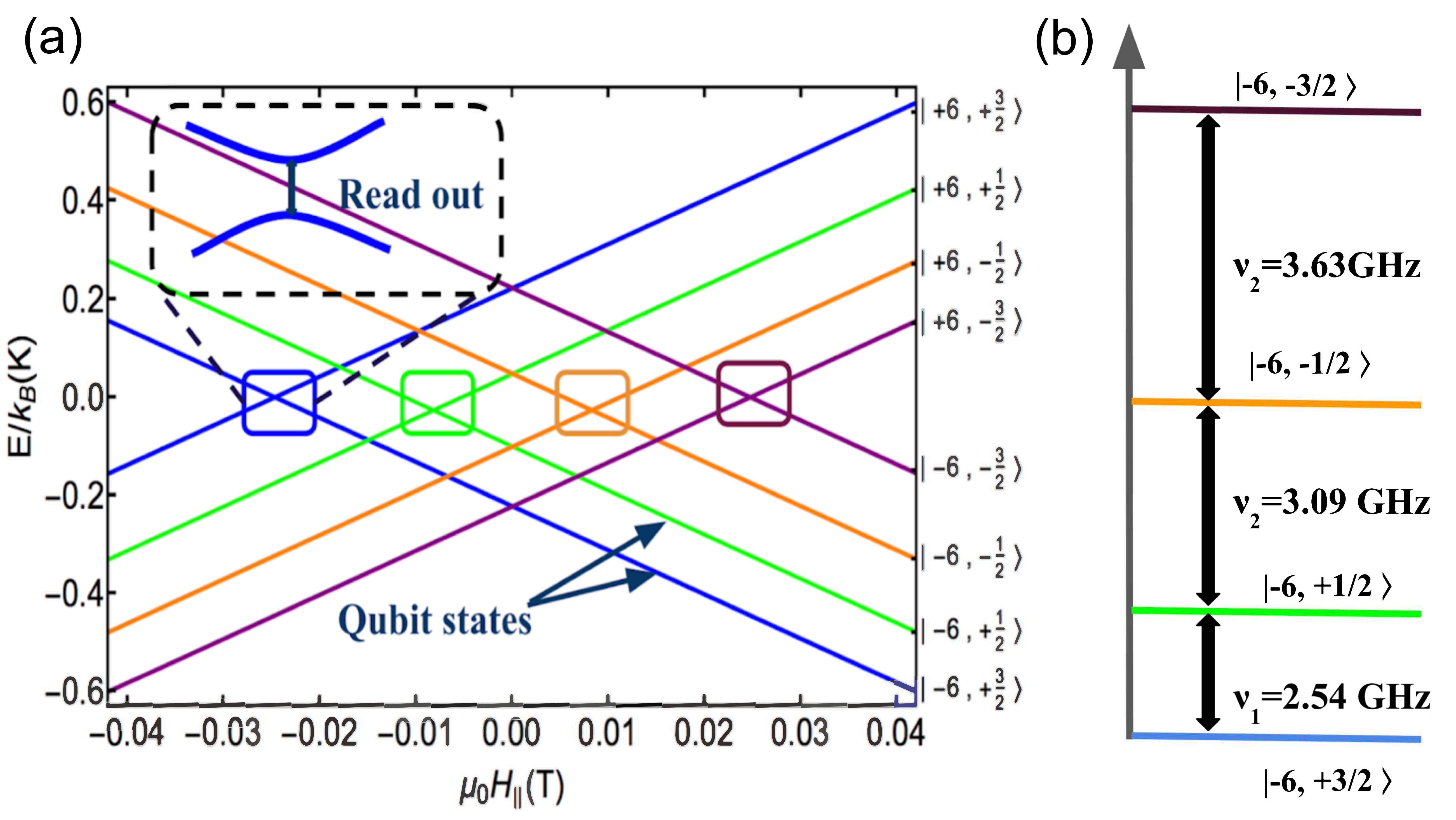}
\caption{(a) Energy level diagram of the lowest eight energy eigenstates of the ${\rm TbPc_{2}}$ molecule. States are labeled by the electronic and nuclear spin quantum numbers of the non-interacting states with which they have greatest overlap. The avoided crossings (boxes) are used to initialize and readout nuclear spin states, while gate operations are performed after tuning away from avoided crossings. One possible choice for the two qubit states is indicated, although this choice is not unique. (b) Zoom-in of four of the levels with energy spacings away from avoided crossings indicated.} 
\label{fig:energy_smm1}
\end{figure}

The TbPc$_2$ molecule consists of a $\rm Tb^{3+}$ ion sandwiched between two flat ${\rm Pc}$ ligands (see Fig~\ref{fig:smm_resonator}).\cite{Ishikawa2003,Thiele_Science14,Wernsdorfer_Nature,Wernsdorfer_Nanotech,Wernsdorfer_views,ROBA15} The  $\rm Tb^{3+}$ ion has an electronic configuration of $\rm [Xe]4f^8$, which implies a total orbital angular momentum of $L=3$ and a total spin of $S=3$ based on Hund's rules. Therefore, the electronic ground state has total angular momentum $J=6$. An unusually strong spin-orbit coupling ($\sim$2900 K) combined with the ligand field lifts the degeneracy of the $J=6$ multiplet and separates the ground state doublet $m_J=\pm 6$ from the first excited state doublet $m_J=\pm 5$ by an energy gap of 600 K (zero-field splitting). Consequently, at very low temperatures ($\sim$50 mK), the ground states with $m_J=\pm 6$ become well isolated, and the electronic spin becomes Ising-like.  In describing the energy units in Kelvin we have set $\hbar=k_b=1$.
Furthermore, the $\rm Tb^{3+}$ ion contains a nuclear spin of $I=3/2$ that couples to the electronic spin via a hyperfine interaction. This hyperfine coupling (with strength $A=24.9$ mK~\cite{Thiele_Science14}) lifts the four-fold degeneracy of the nuclear spin, yielding a low-energy manifold of eight non-degenerate electron-nuclear spin states as shown in Fig.~\ref{fig:energy_smm1}. The hyperfine interaction also contains a quadrupolar term (with coupling strength $P=0.4$ mK~\cite{Thiele_Science14}) which results in the non-uniform energy spacing evident in the figure. In addition, the off-diagonal part of the ligand field couples the electronic states $m_J=\pm 6$ and thus creates avoided crossings on the order of $1{\rm \mu K}$ between states with the same nuclear spin projection (marked with boxes in Fig.~\ref{fig:energy_smm1}). These avoided crossings are used to initialize and readout the nuclear spin states through quantum tunneling of magnetization.~\cite{Thiele_Science14} 

The properties summarized above are captured by the following effective Hamiltonian:
\begin{eqnarray}\label{Ham_smm}\
H_{SMM}=H_{Z}+H_{LF}+H_{HF},
\end{eqnarray} 
which includes contributions from Zeeman interactions, the ligand field, and the hyperfine interaction:
\begin{eqnarray}\label{Ham_zeeman}\
H_{Z}=g_{l}\mu_{B}\mathbf{J}\cdot\mathbf{B},
\end{eqnarray} 
\begin{eqnarray}\label{Ham_LF}\
H_{LF}=H_{LF}^{D}+H_{LF}^{OD},
\end{eqnarray} 
\begin{eqnarray}\label{Ham_hf}\
H_{HF}&=&A\mathbf{I}\cdot\mathbf{J}+P\left[I_{z}^{2}-\frac{1}{3}(I+1)I\right],
\end{eqnarray} 
where, $g_{l}=1.5$, $\mu_{B}$ is the Bohr magneton, and the ligand field is described in Ref.\cite{Thiele_thesis}. The energy levels shown in Fig.~\ref{fig:energy_smm1} are the lowest eight eigenstates of this Hamiltonian plotted as a function of the magnitude of the external magnetic field, which is chosen to point along the $z$ direction. It is clear that far away from the avoided crossings, the energy splittings between states with the same $m_J$ are approximately constant. Restricting attention to the $m_J=-6$ submanifold, we have after diagonalization
\begin{eqnarray}\label{Ham_diag}\
H_{SMM}^{D}= \sum_{j=1}^{4}\omega_{j}|j\rangle \langle j|.
\end{eqnarray}
Away from the avoided crossings, the eigenenergies $\omega_j$ depend approximately linearly on the magnetic field, with splittings given by (in GHz) $\nu_{1}=2.54$, $\nu_{2}=3.09$, and $\nu_{3}=3.63$ (see Fig.~\ref{fig:energy_smm1}). 

Most nuclear spin qubit proposals make use of time-dependent magnetic fields to manipulate the spin states.\cite{Kane_Nature98,Pla_Nature13,Laucht_SciAdv15} Although very high single-qubit gate fidelities have been achieved with this approach,\cite{Pla_Nature13,Laucht_SciAdv15} the speed of the gates is limited by restrictions on the amplitude of magnetic pulses. In order to avoid substantial cross-talk and joule heating caused by the micro-coil used to generate the magnetic field, the amplitude must typically be kept below a few mT.\cite{Thiele_thesis} To overcome this problem, the manipulation of the nuclear spin state by means of electric fields has been proposed for TbPc$_2$ SMMs\cite{Thiele_Science14} and phosphorous donors in silicon.\cite{Sigillito_NatNano17,Asaad_arxiv} Since the electric field does not directly couple to spin, it is necessary to have an intermediate interaction which converts an ac electric field into an effective magnetic field. Various mechanisms have been used to facilitate this conversion, including through spin-orbit coupling,\cite{Nowack_Science07} magnetic field gradients,\cite{Takeda_SciAdv16,Mi_Science18,Samkharadze_Science18} and hyperfine interactions.\cite{Thiele_Science14,Sigillito_NatNano17} Here, we focus on the hyperfine Stark effect, which was experimentally demonstrated to yield electrically driven Rabi oscillations between nuclear spin states in TbPc$_2$.\cite{Thiele_Science14}

The hyperfine Stark effect refers to the shift in nuclear spin energy levels caused by an applied electric field. This effect originates from the dependence of the hyperfine couplings on the shape of the electronic wavefunction, which is of course sensitive to electric fields. We can rewrite the hyperfine Hamiltonian as an effective Zeeman interaction, $H_{hf}=g_N\mu_N \mathbf{I}\cdot\mathbf{B}_{eff}(A,J)$, where $\mathbf{B}_{eff}$ is an effective magnetic field felt by the nuclear spin due to a net electronic spin magnetization. By substituting $J=6$ and $g_N=1.354$, we get $B_{eff}=313$ T, showing that the effective magnetic field created by an ac electric field is several orders of magnitude larger than the actual magnetic fields produced by micro-coils.\cite{Thiele_Science14} In Ref. ~\cite{Thiele_Science14} it was found from both experimental results and perturbation theory calculations that the sensitivity of the hyperfine coupling to an applied electric field $E$ is approximately given by $\Delta A/A\sim10^{-3}$ for fields on the order of $E\sim 1$mV/nm.

Although the experimental demonstrations of Ref.~\cite{Thiele_Science14} make it clear that electrically driven nuclear spin transitions are enabled by a significant hyperfine Stark effect in this system, many details have yet to be clarified. Most importantly, the precise form of the hyperfine tensor for TbPc$_2$ is not yet known, giving rise to uncertainty in precisely how the nuclear spin states respond to electric fields. This issue is critical not only for improving the quality of single-qubit operations, but also for designing schemes to couple multiple qubits together via electrical interactions. Here, we shed some light on the nature of the hyperfine interaction by determining the simplest hyperfine tensor necessary to produce Rabi oscillations.

To investigate this matter, we start with the most general form of the hyperfine interaction 
\begin{eqnarray}\label{Ham_hf_gen}\
H_{HF}=\sum_{\alpha\beta}\textbf{I}_{\alpha}A_{\alpha\beta}\textbf{J}_{\beta},
\end{eqnarray} 
where $A_{\alpha\beta}$ is a matrix representing the (generally anisotropic) coupling of the electronic and nuclear spins. An applied electric field will shift the hyperfine interaction, which to first order in the field yields a second term of the same form:
\begin{eqnarray}\label{Ham_hf_gen}\
H_{HF}\approx\sum_{\alpha\beta}\textbf{I}_{\alpha}A_{\alpha\beta}\textbf{J}_{\beta}+\alpha E(t)\sum_{\alpha\beta}\textbf{I}_{\alpha}A_{\alpha\beta}\textbf{J}_{\beta},\label{eq:perturbedHF}
\end{eqnarray} 
where $E(t)$ is the electric field, and $\alpha$ is a constant. Our qubit states are defined to be the lowest energy eigenstates of $H_{SMM}$, which includes the first term in Eq.~\ref{eq:perturbedHF} but not the second. The second term allows us to drive transitions between the different energy eigenstates, and we thus refer to it as the control Hamiltonian, $H_c(t)$. Here, the time dependence reflects that of the applied electric field. We see that the controllability of the TbPc$_2$ nuclear spin qubit is determined by the matrix elements of $H_{HF}$ with respect to the lowest-energy eigenstates of the full Hamiltonian $H_{SMM}$. We find that all of these matrix elements, taken with respect to the states depicted in Fig.~\ref{fig:energy_smm1}, vanish identically if $A_{\alpha\beta}$ is purely diagonal. Thus, in order to drive Rabi oscillations between states in the low-energy manifold, it must be the case that at least one off-diagonal entry of $A_{\alpha\beta}$ is nonzero. Furthermore, we want to choose the $z$ axis to be along the easy anisotropy axis so that the ground state doublet is $J_z=+6$ and $J_z=-6$. Thus, we consider the simplest case where only one off-diagonal component is nonzero, and we take this to be $A_{xz}$. Given that Rabi oscillations have been demonstrated experimentally, we know that such a term must be present. The presence of $A_{xz}$ or $A_{yz}$ terms reflects a deviation from the 4-fold axis symmetry\cite{Ishikawa2005}. This can be caused by the transverse electric field or by deviations of the molecular structure from the ideal $D_{4h}$ symmetry.~\cite{Alex_arxiv} Determining the precise
nature of this anisotropy requires detailed ab-initio
calculations that we leave to future work.

Taking the diagonal entries of $A_{\alpha\beta}$ to be the same for simplicity (all equal to $A$) and retaining only $A_{xz}$ from the off-diagonal entries, we arrive at the following form for the control Hamiltonian,
\begin{eqnarray}\label{Ham_hf_gen}\
H_{c}(t)=\eta A \cos(\omega_p t) \mathbf{n}\cdot\mathbf{I},\label{eq:controlham}
\end{eqnarray} 
where $\mathbf{n}=\sin(\theta)\hat{z}+\cos(\theta)\hat{x}$, $\theta=\arctan{A/A_{xz}}$, $\omega_p$ is the frequency of the oscillating electric field, and $\eta$ is a constant that depends on $\alpha$, $J$, $m_J$, and the magnitude of the electric field. To arrive at Eq.~\ref{eq:controlham}, we have projected the electronic angular momentum $\mathbf{J}$ onto the $m_J=-6$ submanifold since our focus will be on driving transitions between states within this manifold. 

\begin{figure}[h]
  \centering
  \includegraphics[width=0.45\columnwidth]{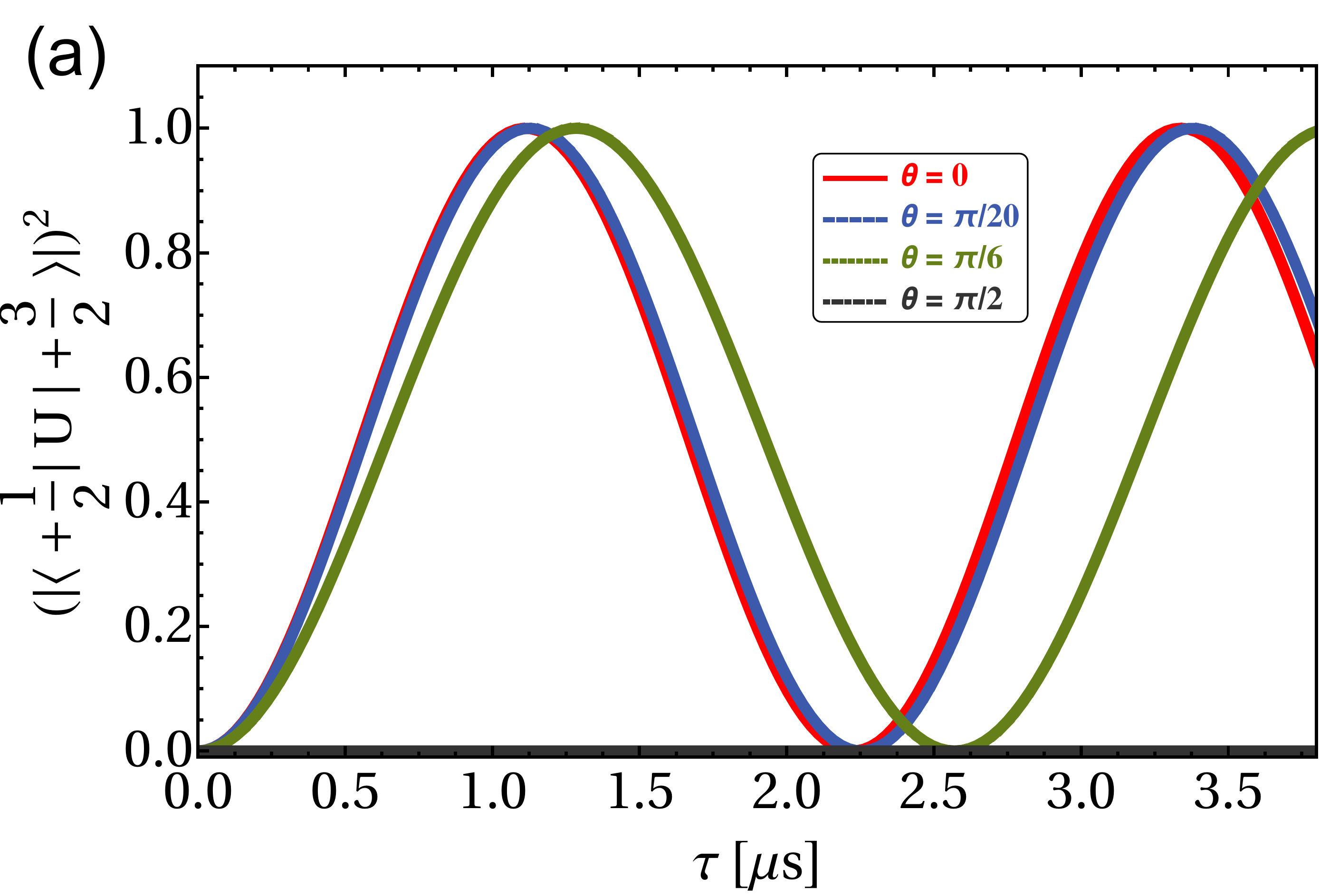}
    \includegraphics[width=0.45\columnwidth]{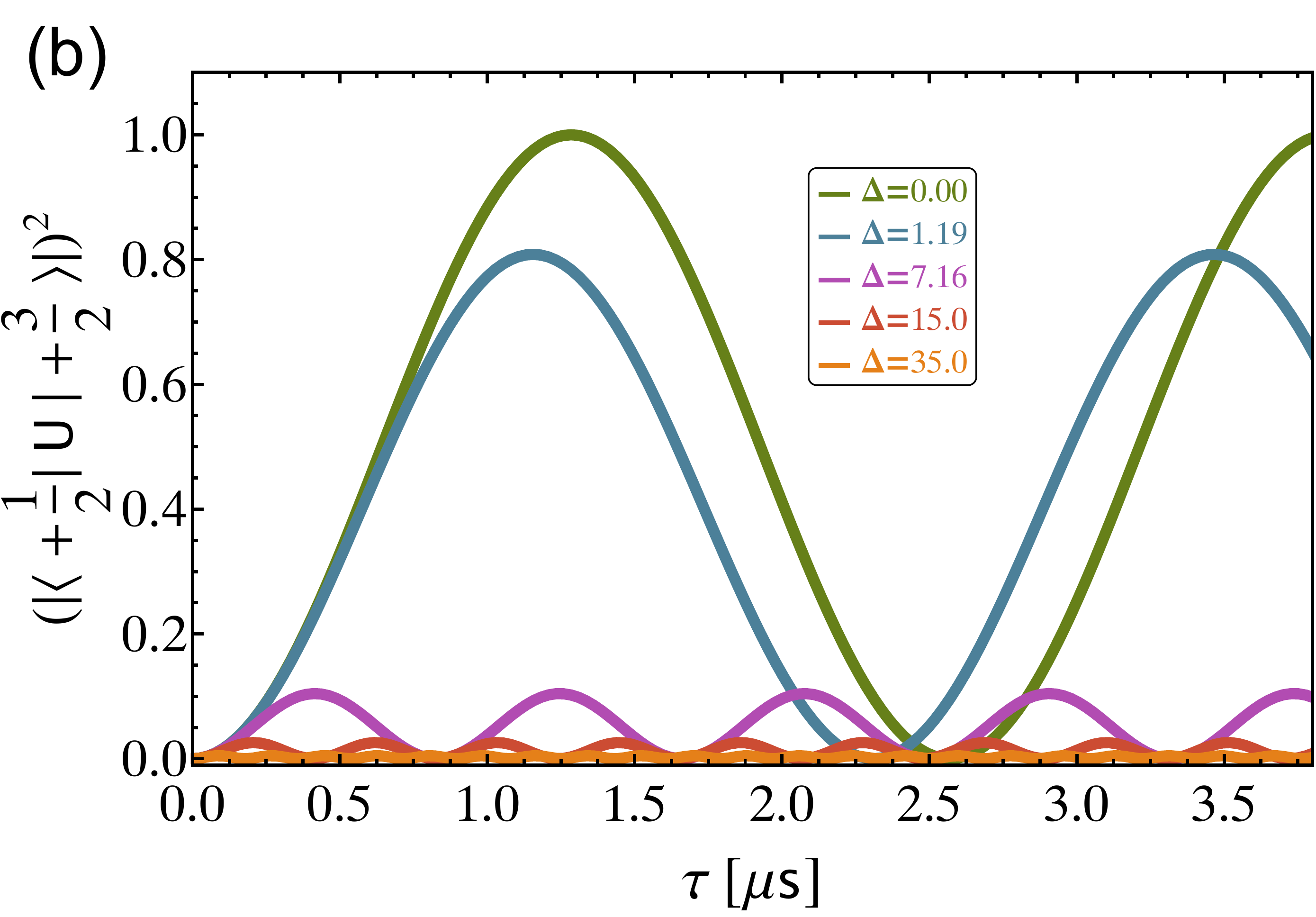}
\caption{Rabi oscillations. An ac electric field drives transitions between the lowest two nuclear spin states $|-6,+3/2\rangle$ and $|-6,+1/2\rangle$. (a) The transition probability as a function of driving time for several different values of the hyperfine anisotropy parameter $\theta$. The Rabi periods obtained from the formula in the main text are $T_R[\theta=0]=2.23$ $\mu s$ and $T_R[\theta=\pi/6]=2.57$ $\mu s$, which agree well with numerical results. (b) The transition probability as a function of driving time for five different values of the detuning. The values of detuning $\Delta$ are in MHz, and we have set $\theta=\pi/6$. We set $\eta=0.001$ in both panels.} 
    \label{fig:Rabi}
\end{figure}

Now that we have established a form for the control Hamiltonian, we proceed to investigate the controllability of the TbPc$_2$ qubit as a function of the hyperfine anisotropy parameter $\theta$. We focus on the lowest energy states $|\!-\!6,+3/2\rangle$ and $|\!-\!6,+1/2\rangle$ as our qubit states, which are separated in energy by $\nu_1=2.54$ GHz, although the same analysis could be applied for any two nuclear spin states. Notice that in this two-level subspace we can effectively make the replacements $I_x\to\sqrt{3}\sigma_x$ and $I_z\to\sigma_z$ where $\sigma_x$ and $\sigma_z$ are Pauli matrices. Initially, we consider the case of resonant driving, for which the detuning vanishes: $\Delta=\omega_p-\nu_1=0$. Solving the time-dependent Schr\"odinger equation for the evolution operator $U$ with the Hamiltonian from Eq.~\ref{eq:controlham}, we obtain the transition probability as a function of time as shown in Fig~\ref{fig:Rabi}. Fig.~\ref{fig:Rabi}(a) shows the resulting Rabi oscillations for several different values of $\theta$. As expected, only the component in the $x$ direction is capable of driving transitions between the states, while the population transfer is zero for $\theta=\pi/2$. Importantly, we see that for any finite amount of anisotropy, it is possible to completely transfer the population from one state to the other. Moreover, while the transfer becomes slower as the anisotropy is reduced, the transfer time increases slowly with increasing $\theta$. This indicates that the performance of single-qubit gates is relatively insensitive to the precise form of the hyperfine tensor.

The fact that Eq.~\ref{eq:controlham} yields a $\sigma_z$ term in addition to $\sigma_x$ makes the present control problem a bit different from the standard Rabi problem. Thus, it is worth checking the extent to which the usual Rabi behavior applies here. For a general detuning $\Delta$, and without ignoring the fast oscillating field in the interaction representation (no rotating wave approximation applied) the Rabi frequency is given by the formula $\Omega_R/2\pi=\sqrt{(\Delta/2\pi)^2+(\sqrt{3}g_{N}\mu_N B_x/h)^2}$, where $B_x$ is the transverse component of the effective magnetic field. When $\Delta=0$, this leads to the following expression for the Rabi period: $T_R=2\pi/\Omega_R=4\pi/(\sqrt{3}\eta A\cos(\theta))$, which agrees well with the numerical results shown in Fig.~\ref{fig:Rabi}(a). We note that our numerical results for the Rabi frequencies are also compatible with the reported experimental values, which are on the order of a few $\mu$s.\cite{Thiele_Science14} Fig.~\ref{fig:Rabi}(b) shows the behavior of the Rabi oscillations for off-resonant driving, $\Delta\ne0$. As is the case for the standard Rabi problem, the Rabi frequency is minimal at resonance and increases as one tunes away from resonance. We conclude that by adjusting the driving time and detuning, it is possible to create any single-qubit gate for which the rotation axis is in the $xz$ plane. All other single-qubit gates can be obtained by concatenating these operations using standard composite pulse sequences.\cite{Goelman_JMR89}

While we have seen that it is possible to perform any single-qubit gate on isolated TbPc$_2$ qubits, this is not guaranteed to remain true when we start coupling two or more qubits together. For example, we need to ensure that it is possible to address each qubit individually without disturbing the rest. This is achievable by taking advantage of the dc stark effect induced by applying a dc gate voltage. Such a voltage will shift the energy levels of the nuclear spin states, allowing us to adjust the qubit resonance frequency at will. Thus, when we perform an operation on one qubit, we can first tune it away from the other qubits to avoid driving them. As we will see later, this ability to shift the resonance frequency is also crucial to achieving high-fidelity entangling gates. In Refs.~\cite{Thiele_Science14,Thiele_thesis}, shifts in the nuclear spin resonance frequency of $\Delta \nu_{1}^{exp}= 1.72$ MHz and $7.03$ MHz were measured and compared with perturbation theory for gate voltages of $V_g=10$ mV and $16$ mV, respectively. These values in turn correspond to shifts of the hyperfine constant on the order of $\Delta A/A=5.6\times 10^{-4}$ and $\Delta A/A=2.3\times 10^{-3}$. To check these findings, we have used the perturbation theory results\cite{Thiele_Science14} where we included the hyperfine constant shifts in our numerical simulation and computed the resulting frequency shifts in each case, which yielded slightly
different numerical values: $\Delta \nu_{1}^{num}= 1.19$ MHz and $\Delta \nu_{1}^{num}=7.16$ MHz for $V_g=10$ mV and $16$ mV, respectively. As we can see from Fig~\ref{fig:Rabi}(b), the larger voltage offsets should be sufficient to decouple the qubit from the driving field. Next, we use both the experimental and numerical results for the frequency shifts to estimate the coupling strength between a TbPc$_2$ qubit and the electric field of a microwave resonator and to design high-fidelity two-qubit entangling gates.

Entangling gates are a requirement for any universal quantum computer and a basic ingredient for all quantum algorithms of interest. Creating entanglement on demand requires sufficiently strong, controllable interactions between qubits. However, this is notoriously difficult to achieve for qubits based on the spin of a donor atom or molecule. This is because the two main options for spin-spin coupling, namely dipolar couplings and exchange interactions, are either too weak or diminish too quickly with distance and thus require the capability to place the spins in close proximity to each other with high accuracy. The latter is due to the strong confinement of the electronic wavefunction around the donor or molecule. Longer-range spin-spin couplings mediated by resonators have been proposed previously,\cite{Childress_PRA04,Imamoglu_PRL09,Schuster_PRL10,Luis2016} but these schemes are normally based on magnetic interactions that are again too weak (10-100 Hz) to be practical for achieving coherent interactions between individual spins. To overcome this issue, we can instead consider using the electric field of a superconducting transmission line resonator to mediate interactions between TbPc$_2$ qubits. While this approach was originally developed to couple superconducting qubits,\cite{Wallraff_Nature04,Majer_Nature07} recently, it has been successfully implemented to create long-distance coupling between electron spins in semiconductor quantum dots\cite{Borjans_arxiv19} and between electron spins and superconducting qubits.\cite{Landig_arxiv19} A similar approach has also been proposed for qubits based on the nuclear spin of phosphorous donors in silicon.\cite{Tosi_PRB18}

 To obtain a stronger coupling between a TbPc$_2$ SMM and a resonator, we can again leverage the hyperfine Stark effect to couple the molecule to the electric field of the cavity instead of its magnetic field. For this purpose, resonators based on NbTiN nanowires are particularly promising. These are microwave-frequency resonators that possess a large kinetic inductance, and they have already been successfully coupled to spins in semiconductor quantum dots.\cite{Borjans_arxiv19,Burkard_arxiv19,Landig_arxiv19} There are several reasons for choosing this particular type of resonator. First of all, it has a high critical magnetic field (B$\sim$350 mT) that is well above the fields used in TbPc$_2$ experiments.\cite{Samkharadze_physrevappl} Second, the high kinetic inductance of the nanowires leads to an increase in the characteristic impedance up to $Z_r\sim$1 k $\hbar{\Omega}$, which is two orders of magnitude larger than what is typically achieved in coplanar waveguides. Furthermore, the increase in impedance leads to an increased resonator vacuum rms voltage of $V_{RMS}\sim$20 $\mu$V, which in turn produces larger shifts in the TbPc$_2$ hyperfine coupling.\cite{Samkharadze_physrevappl} We now give an estimate of the resulting SMM qubit-resonator coupling. Earlier we noted that a dc gate voltage on the order of 10 mV produces a shift in the qubit resonance frequency on the order of 1-2\%., as reported in Ref. 16. There is currently not enough experimental data on how the resonance frequencies depend on gate voltage to extrapolate this finding to other voltages. This dependence factors critically into the effective qubit-resonator coupling, highlighting the need for further experimental work along these lines. To proceed with our estimate of the coupling, we instead rely on the perturbation theory result of Ref. 16, which gives a linear dependence of the hyperfine interaction on the gate voltage, implying that a resonator vacuum rms voltage of 1-20 $\mu$V can produce a frequency shift of up to 0.002\%. Combining this with the value for the hyperfine constant, $A=518$ MHz, we estimate the qubit-resonator coupling to be $g/(2\pi)\sim 2\times10^{-5}|m_J|A\sim60$ kHz. While this value is well above couplings generated by magnetic interactions, it lies below that of other systems where an electrical spin-resonator coupling of order 1-10 MHz has been achieved.\cite{Mi_Science18,Landig_Nature18,Samkharadze_physrevappl} To determine whether this estimate can be considered to lie within the strong coupling regime, we must compare it to the resonator decay rate and the spin dephasing time. Assuming a resonator quality factor of $Q\sim10^5$ and a cavity frequency on the order of $1$-$10$ GHz\cite{Samkharadze_physrevappl}, the cavity decay rate is $\kappa/(2\pi)=\omega_c/Q\simeq 10$ kHz, a little below our estimated qubit-resonator coupling. The spin dephasing rate is given by $\gamma/(2\pi)=1/T_{2}^{*}\simeq3$ kHz, where we have used the measured value of the dephasing time: $T_{2}^{*}\simeq0.3$ ms.\cite{Moreno_CSR18} These numbers suggest that the TbPc$_2$ qubit-resonator system is currently near the edge of the strong coupling regime ($g>\gamma,\kappa)$. It may be possible to increase coherence times further by using better substrates to eliminate sources of noise and by employing dynamical decoupling schemes (taking advantage of the fact that charge noise---the dominant type of noise in this system---is concentrated at low frequencies) to
make $T_{2}\gg T_{2}^{*}$, rather than $T_{2}^{*}$, the relevant timescale. On the other hand, improving the quality factor of resonators much beyond $10^5$ may not be feasible. However, it may be possible to enhance the hyperfine constant itself and/or its sensitivity to electric fields. Both depend on the ligand field, which in turn could likely be influenced by external factors such as the choice of substrate. Determining the extent to which the electrical coupling can be increased first requires a deeper understanding of what determines the hyperfine constant and how the TbPc$_2$ molecule responds to its environment.

To obtain a better understanding of how much stronger the qubit-resonator coupling needs to become, we now investigate the performance of two-qubit entangling gates as a function of the interaction strength. We begin by writing down a Hamiltonian that describes multiple TbPc$_2$ qubits coupled to a common resonator mode:
\begin{eqnarray}\label{Ham_two-qubit}\
H_{0}&=&\omega_ca^{\dagger}a+\sum_{n,j}\omega_{n,j}|n,j\rangle\langle n,j|\nonumber\\&+&(a+a^{\dagger})\sum_{n,j}\epsilon_{n,j}|n,j\rangle\langle n,j|\\
&+&\sum_{n,j}\left(\xi_{n,j}^{-}a^{\dagger}|n,j\rangle\langle n,j+1|+\xi_{n,j}^{+}a|n,j+1\rangle\langle n,j|\right),\nonumber
\end{eqnarray} 
where $\omega_c$ denotes the resonator frequency, $\omega_{n,j}$ indicates the $j$th energy level of qubit $n$ (here we consider $n=1,2$), $\epsilon_{n,j}\equiv\eta A \sin(\theta_{n})$, $\xi_{n,j}^{\pm}\equiv \eta A\cos(\theta_{n})\sqrt{I(I+1)-j(j\pm1)}$, and $I=3/2$ is the total spin of the nucleus. Here, we have made the rotating wave approximation in which we remove counter-rotating terms under the assumption that $\omega_{n,j+1}-\omega_{n,j}\sim\omega_c$. This is essentially a Jaynes-Cummings-type Hamiltonian, but with an additional $\epsilon_{n,j}$ term which implements energy-level tuning in addition to the inter-level transitions generated by the usual Jaynes-Cummings $\xi_{n,j}^{\pm}$ terms. 

In order to perform the maximally entangling two-qubit gates needed for quantum computing algorithms, it is sufficient to electrically drive a single SMM qubit that is resonator-coupled to a second qubit. The most significant source of gate errors in this case is leakage to nuclear spin states outside the logical subspace or to excited resonator states. Our strategy to address this leakage is based on pulse designs using analytical approaches. To implement high-fidelity two-qubit entangling gates, we employ a recently developed formalism known as the SWIPHT protocol.\cite{Economou_PRB15} This method enables a speedup of the two most common entangling gates (CZ and CNOT) that can range from a factor of two to more than one order of magnitude while maintaining high fidelities and using only smooth pulses given by analytical expressions. In this paper, we focus on the well-known two-qubit entangling CNOT gate in which the state of one qubit is flipped conditionally on the state of the other qubit.
 
 Diagonalizing the two-qubit-resonator Hamiltonian given in Eq.~\ref{Ham_two-qubit}, we obtain the interacting dressed states. We define our logical qubit states to be the four dressed states that have the largest overlap with the non-interacting two-qubit states $|00\rangle$, $|01\rangle$, $|10\rangle$, $|11\rangle$. We denote the logical states as $\widetilde{|00\rangle}$, $\widetilde{|01\rangle}$, $\widetilde{|10\rangle}$, $\widetilde{|11\rangle}$. Here, we consider identical qubit-resonator couplings for both qubits, however, our analysis can be adapted straightforwardly to the case of non-identical couplings as well. A microwave electrical pulse drives transitions between the logical states as described by the following control Hamiltonian:
\begin{eqnarray}\label{driving_ham}\
H_p&=&\Omega(t)\cos(\omega_p t)\Big[\lambda_{c} (a+a^\dagger)+\sum_{n,j}\lambda_n(\epsilon_{n,j}|n,j\rangle\langle n,j|\nonumber\\
&+&\xi_{n,j}^{-}|n,j\rangle\langle n,j+1|+\xi_{n,j}^{+}|n,j+1\rangle\langle n,j|)\Big],
\end{eqnarray} 
where $\Omega(t)$ is the amplitude of the pulse and $\omega_p$ its frequency. We included the parameters $\lambda_c$, $\lambda_n$ to indicate which components of the tripartite system are driven by the pulse. Here, we consider the case with $\lambda_c=\lambda_1=0$, $\lambda_2=1$, which means we are only driving the second qubit. 

We can implement a CNOT gate by performing a $\pi$-rotation on the target transition $\widetilde{|00\rangle}\leftrightarrow\widetilde{|01\rangle}$ while avoiding all other transitions in the spectrum. For qubit-resonator couplings that are not too strong, there is typically only one other driven transition that is nearby in frequency: the transition $\widetilde{|10\rangle}\leftrightarrow\widetilde{|11\rangle}$. In the absence of a qubit-resonator coupling, this unwanted transition would be degenerate with the target transition since they both correspond to driving the second qubit between its two states, and without the inter-qubit coupling, this would not depend on the state of the first qubit. When the qubit-resonator coupling is switched on, the two transitions remain nearly degenerate. Typically one would need to resort to long, spectrally selective pulses to avoid exciting this unwanted transition, which leads to slow gates. However, the SWIPHT formalism allows one to avoid long pulses by letting the pulse drive the unwanted transition also, in such a way that it undergoes cyclic evolution and acquires a trivial $2\pi$ phase, so that a CNOT gate is still achieved. It was shown in Ref.~\cite{Economou_PRB15} that pulses which perform a SWIPHT-based CNOT gate can be constructed using a systematic recipe. This approach is based on the fact that, for a driven two-level system, both the driving field $\Omega(t)$ and the time evolution operator $\mathcal{U}(t)$ can be expressed in terms of a single real function $\chi(t)$:\cite{Barnes_PRA13}
\begin{eqnarray}\label{omega_t}\
\Omega(t)=\frac{\ddot{\chi}}{2\sqrt{\frac{\delta^2}{4}-\dot{\chi}^2}}-\sqrt{\frac{\delta^2}{4}-\dot{\chi}^2}\cot(2\chi),
\end{eqnarray} 
\begin{equation}\label{Ut}
\mathcal{U}(t)=e^{-i\frac{\pi}{4}\sigma_y}
\begin{bmatrix}
    \cos\chi e^{i\psi^{-}}       & \sin\chi e^{-i\psi^{+}}  \\
 -\sin\chi e^{i\psi^{+}}     &\cos\chi e^{-i\psi^{-}}  \\
\end{bmatrix},
\end{equation} 
where $\psi^{\pm}=\int^{t}_{0}dt'\sqrt{\frac{\delta^2}{4}-\dot{\chi}^2}\csc{[2\chi(t')]}\pm\frac{1}{2}\arcsin(\frac{2\dot\chi}{\delta})$, and $\delta$ is the pulse detuning. In order to have a valid solution, $\chi$ must satisfy the constraint $|\dot{\chi}|\le |\frac{\delta}{2}|$ along with the initial conditions  $\chi(0)=\pi/4$, $\dot\chi(0)=0$. This formalism has been used to design two-qubit entangling gates in superconducting qubits and quantum dots\cite{Economou_PRB15,Deng_PRB17,CalderonVargas_arxiv19} with fidelities exceeding 99\%. This method has also been experimentally demonstrated in the case of superconducting qubits.\cite{Premaratne_PRA19}

We can use this construction to create a CNOT gate on two SMM qubits by taking the two-level system to be the two states of the second qubit. In this case, we can impose that its evolution be cyclic by requiring that $\chi(\tau)=\pi/4$ and $\dot\chi(\tau)=0$, where $\tau$ is the duration of the pulse. At the same time, we also need to ensure that the pulse performs a $\pi$ rotation on the first qubit, as required for a CNOT. If we take the pulse to be resonant with this qubit, then this is tantamount to requiring that the area of the pulse be equal to $\pi/2$: $\int_{0}^{\tau}dt\Omega(t)=\pi/2$. The following ansatz for $\chi(t)$ can be used to satisfy all these criteria:\cite{Economou_PRB15}
\begin{eqnarray}\label{chi_t}
\chi(t)=C(t/\tau)^{4}(1-t/\tau)^{4}+\pi/4.
\end{eqnarray} 
This ansatz automatically obeys the initial and final conditions on $\chi$. Moreover, we can tune the parameters $C$ and $\tau$ until the pulse area constraint is also satisfied.
We find numerically that the values $C=138.9$ and $\tau=5.87/|\delta|$ achieve this.\cite{Economou_PRB15} Note that since the pulse is resonant with the first qubit, $\delta$ is equal to the difference in resonance frequencies of the target and unwanted transitions. This frequency difference determines the gate time of the SWIPHT pulse, as is clear from the above formula for $\tau$.

To evaluate the performance of the resulting CNOT gate, we numerically solved the time-dependent Schr\"odinger equation in the interaction picture defined with respect to $H_0$ to obtain the evolution operator for the two-qubit system. We define our gate in the interaction picture so that it is created purely from the applied control pulse and does not include additional phases coming from free evolution. We performed this calculation using the pulse obtained from Eqs.~\ref{omega_t} and \ref{chi_t} and for a range of coupling strengths $g$ and resonator frequencies $\omega_c$. In each case, we computed the fidelity of the gate using the standard formula\cite{Pedersen_PLA07} $F\equiv\frac{1}{20}({\rm Tr} [\mathcal{U}\mathcal{U}^{\dagger}]+|{\rm Tr} [\mathcal{U}^{\dagger} \hbox{CNOT}]|^2)$, optimized over single-qubit gates on both qubits. Our simulations include a frequency shift of $40$ MHz on the second qubit, which can be obtained by applying a dc gate voltage on the order $V=90$ mV. We found that this produces better performance in terms of both fidelity and gate speed.

\begin{figure} [httb!]
\includegraphics[width=\columnwidth]{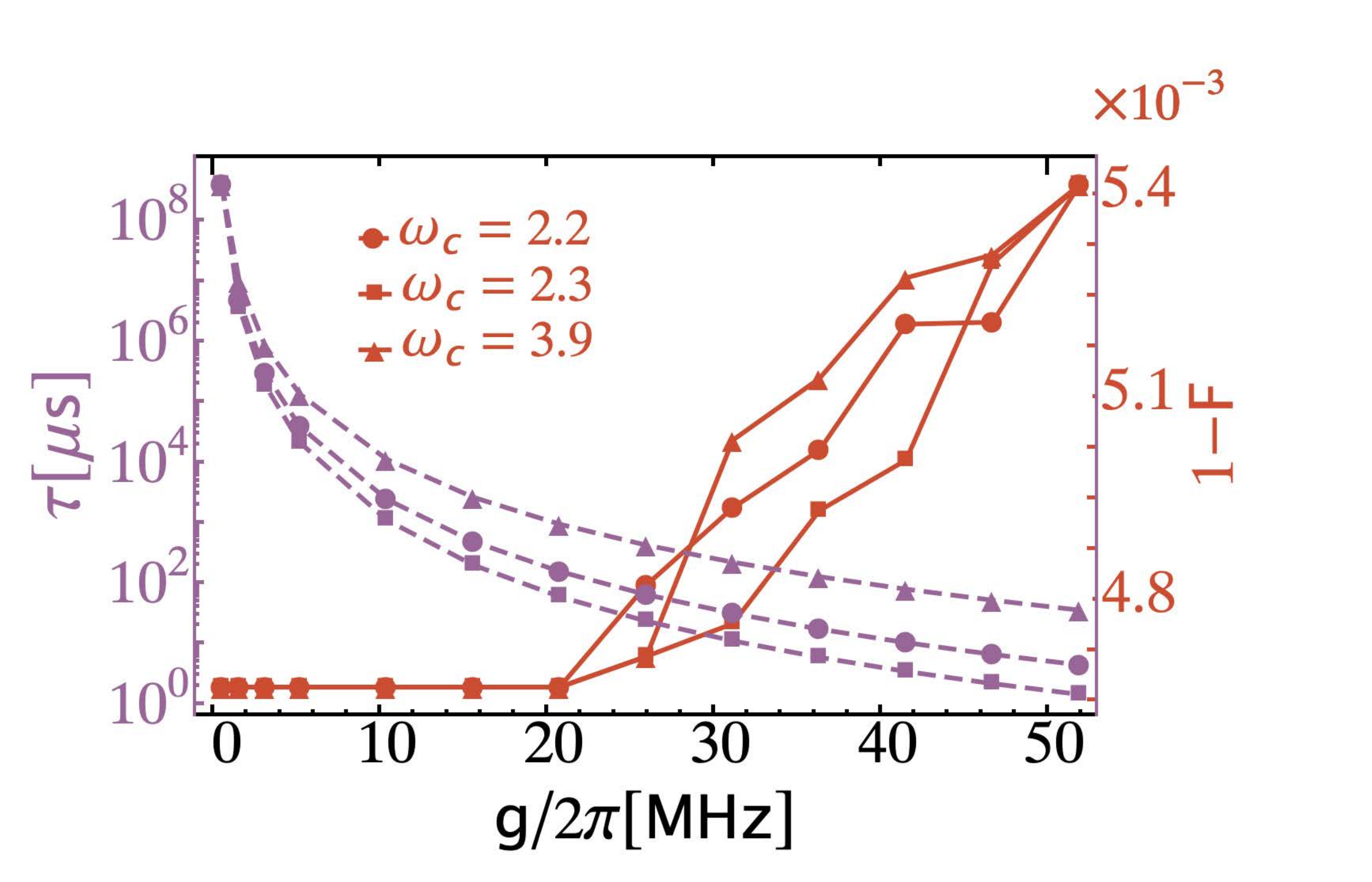}
\caption{Infidelity $1-F$ and gate time $\tau$ of the two-qubit entangling CNOT gate as functions of the qubit-resonator coupling $g$ for three different values of the resonator frequency $\omega_c$ (in GHz). The gate was generated by the voltage pulse defined by Eqs.~\ref{omega_t} and \ref{chi_t}. Other system parameters were chosen as in Fig.~\ref{fig:Rabi}.}
\label{fig:2qubitgate}
\end{figure}

The results are summarized in Fig.~\ref{fig:2qubitgate}, which shows the infidelity $1-F$ and the gate time as a function of the coupling $g$ and for three different resonator frequencies. First, it is evident that the CNOT fidelity remains above 99\% and is largely insensitive to the qubit-resonator coupling over the full range of couplings considered. In fact, the fidelity remains essentially constant for couplings below 20 MHz. This is true for all three resonator frequencies we considered. In contrast, the gate time is very sensitive to the coupling strength: for couplings in the range 40-50 MHz, the gate times are on the order of a few $\mu s$, while for couplings on the order of a few MHz, the gate times approach milliseconds to seconds. Furthermore, we notice that as the resonator frequency is tuned further from the qubit frequencies, the gate time increases further. In the case where the resonator frequency is closest to the qubits, $\omega_c=2.3$ GHz, the coupling would need to be at least $15$ MHz to get the gate time below the dephasing time of $T_2^*\sim0.3$ ms. Although alternative gate designs such as the cross-resonance gate\cite{Paraoanu_PRB06} may lead to shorter gate times, this result highlights the importance of finding ways to further enhance the coupling between TbPc$_2$ qubits and microwave resonators.

In conclusion, we investigated the possibility of using superconducting resonators to achieve strong coupling between TbPc$_2$ nuclear spin qubits by leveraging the hyperfine Stark effect. To better understand the nature of this effect, we examined single-qubit Rabi oscillations, where we found that anisotropy in the electron-nuclear hyperfine interaction is necessary to electrically drive transitions between the nuclear spin states. This anisotropy must be present since such transitions have been demonstrated experimentally. With this result, we then estimated the qubit-resonator interaction, finding that it lies close to the edge of the strong coupling regime. To understand the implications for entanglement creation, we constructed a Hamiltonian that describes two TbPc$_2$ qubits coupled by a resonator and showed that it is possible to perform high-fidelity two-qubit entangling gates with this architecture. However, we find that in order to reduce gate times sufficiently, it may be necessary to increase the qubit-resonator coupling through improved device designs.

\begin{acknowledgement}

K. N. would like to thank George Barron and Fernando Calderon-Vargas for helpful discussions. This work was supported by DOE grant no. DE-SC0018326.

\end{acknowledgement}

\begin{suppinfo}


 The following files are available free of charge.
 \begin{itemize}
   \item  Hamiltonian of TbPc$_2$ molecule, nuclear state of TbPc$_2$ molecule as qudit, coherent manipulution of nuclear state
   
 \end{itemize}

\end{suppinfo}


\providecommand{\latin}[1]{#1}
\makeatletter
\providecommand{\doi}
  {\begingroup\let\do\@makeother\dospecials
  \catcode`\{=1 \catcode`\}=2 \doi@aux}
\providecommand{\doi@aux}[1]{\endgroup\texttt{#1}}
\makeatother
\providecommand*\mcitethebibliography{\thebibliography}
\csname @ifundefined\endcsname{endmcitethebibliography}
  {\let\endmcitethebibliography\endthebibliography}{}

\end{document}